\documentclass[aps,preprint]{revtex4}
\usepackage[dvips]{graphicx}
\usepackage{amsmath}
\usepackage{color}

\begin{document}
\draft
\title{\bf Experimental investigation of distributions of the off-diagonal elements of the scattering and the Wigner's
$\hat K$ matrices for networks with broken time reversal invariance}
\author{Micha{\l} {\L}awniczak,$^{1}$ Bart van Tiggelen,$^{2}$ and Leszek Sirko$^{1}$}
\address{$^{1}$Institute of Physics, Polish Academy of Sciences,
Aleja \ Lotnik\'{o}w 32/46, 02-668 Warsaw, Poland \\
$^{2}$University Grenoble Alpes, CNRS, LPMMC, 38000 Grenoble, France
}
\date{\today}

\bigskip

\begin{abstract}

We present an extensive experimental study of the
distributions of the real  and imaginary parts of the off-diagonal elements of the scattering matrix $\hat S$ and
the Wigner's reaction $\hat K$-matrix for open microwave networks with broken time ($T$) reversal invariance. Microwave Faraday circulators were applied in order to break $T$-invariance.
The experimental distributions  of the real  and imaginary parts of the off-diagonal entries of the scattering matrix  $\hat S$  are compared with the theoretical predictions from the supersymmetry random matrix theory [A. Nock, S. Kumar, H.-J. Sommers, and T. Guhr, Annals of Physics {\bf 342}, 103-132 (2014)]. Furthermore, we show  that the experimental results are in very good agreement with the recent  predictions for the
distributions of the real  and imaginary parts of the off-diagonal elements of
the Wigner's reaction $\hat K$-matrix obtained within the framework of the Gaussian unitary ensemble of random matrix
theory (RMT) [S. B. Fedeli and Y. V. Fyodorov, J. Phys. A: Math. Theor. {\bf 53}, 165701 (2020)].
Both theories include losses as tunable parameters and are therefore well adapted to the experimental verification.

\end{abstract}

\pacs{05.45.Mt,03.65.Nk}
\bigskip
\maketitle

\smallskip

Quantum chaotic scattering was introduced  almost seventy years ago  to describe  properties of large scale complicated quantum systems \cite{Wigner1951,Haake2001,Weidenmuller2009}. The significant role of broken time-reversal in quantum chaos \cite{Berry1985} is undoubtedly one of the most remarkable discoveries.  It is generally acknowledged that controllable experimental investigations of complex quantum systems are difficult to perform due to decoherence, therefore,  the multitude of physical problems from the field of quantum chaos can be tackled experimentally with the help of microwave networks simulating quantum graphs \cite{Hul2004,Lawniczak2010,Lawniczak2019b}.

This article demonstrates how microwave networks can be applied to obtain the first experimental results on the distributions of off-diagonal elements of the scattering matrix $\hat S$ and the Wigner's reaction  $\hat K$-matrix  for systems with broken time-reversal invariance. The experimental results are compared to the recent exact RMT solutions of these problems \cite{Guhr2014,Fyodorov2020}.

Quantum graphs consisting of one-dimensional wires connected by vertices were introduced
by Linus Pauling \cite{Pauling}. They can be considered as idealizations of physical networks in the limit where the widths of the wires are much smaller than their lengths \cite{Kottos1997}. Quantum graphs constitute rich tools for the study of open quantum systems which exhibit chaotic scattering \cite{Kottos2000,Lawniczak2008,Pluhar2014}.

 They have been used to describe a large variety of systems and models, e.g.,  quantum circuits in tunnel junctions \cite{Namarvar2016}, superconducting quantum circuits \cite{Jooya2016}, realization of high-dimensional multipartite quantum states \cite{Krenn2017} and  discrete-time models of quantum gravity  \cite{Arrghi2017}.

Quantum graphs can be simulated by microwave networks because there is a direct analogy between the Schr\"odinger equation  applied to a quantum graph and the telegraph equation of the corresponding microwave network \cite{Hul2004,Lawniczak2010,Lawniczak2019b}. Microwave networks can be specially designed to allow for the experimental simulation of quantum systems corresponding to all three fundamental ensembles in the  random matrix theory: the Gaussian orthogonal ensemble (GOE, symmetry index $\beta=1$ in RMT) \cite{Hul2004,Lawniczak2008,Hul2012,Sirko2016,Dietz2017,Lawniczak2019} and the Gaussian symplectic ensemble (GSE, symmetry index $\beta=4$) \cite{Stockmann2016} which are both characterized by $T$-invariance, as well as  the Gaussian unitary ensemble (GUE, symmetry index $\beta=2$) \cite{Hul2004,Lawniczak2010,Lawniczak2019b,Bialous2016} for which T-invariance is broken.

The prediction of sypersymmetry for the distribution of off-diagonal entries of the scattering matrix $\hat S$ for systems with losses  was given in Refs. \cite{Guhr2014,Dietz2013} and compared  with data from $T$-invariant microwave cavities \cite{Dietz2013}. The theoretical prediction for the full statistics of the off-diagonal cross sections for $T$-invariant and  $T$-violated systems were calculated exactly by Kumar {\it et al} \cite{Guhr2017}  but the confrontation to data could only be done for a microwave cavity and a compound nucleus that has  T-invariance obeyed.
The first study of  off-diagonal elements of the scattering matrix in open cavities with broken T-invariance
 was done by Dietz {\it et al}   \cite{Dietz2019},  where the distributions of the modulus of the measured off-diagonal
$\hat S$-matrix element $S_{ba}$ for a flat superconducting microwave billiard with weakly broken time-reversal $T$-invariance were compared to RMT simulations.

Although chaotic open systems with  $T$-invariance (GOE systems) have been investigated in most of their aspects,
the theoretical investigations of the Wigner's reaction $\hat K$-matrix were initially concentrated on the  distributions $P(v)$ and  $P(u)$ of the imaginary  and the real parts  of its diagonal elements \cite{Fyodorov2004,Savin2005}. The Wigner's reaction $\hat K$-matrix is a very important observable operator of an open system because it links the properties of {\it closed} chaotic systems, described by a Hamiltonian exhibiting quantum chaos,  to the properties of the corresponding {\it open}  system where irreversible scattering occurs towards the environment outside. The theoretical findings \cite{Fyodorov2004,Savin2005} were confirmed with very good precision  in the experiments using microwave networks \cite{Lawniczak2008,Hul2005} and microwave cavities \cite{Anlage2005,Anlage2006}.

Recently the first experimental results were also reported for the GUE systems \cite{Lawniczak2019b}. In this case the distributions of the diagonal elements of the  $2 \times 2$ Wigner's reaction matrix $\hat K$ were investigated for the large parameter $\gamma=2\pi \Gamma /\Delta \geq 19.4$,  where $\Gamma$ and $\Delta$ are the average width of resonances caused  by both absorption and leaks and the mean level spacing, respectively. Since $\Gamma $ depends on absorption, it is a parameter that is relatively easy to control experimentally with microwave networks.

In the recent paper \cite{Fyodorov2020} the off-diagonal entries $K_{ab}$ of the Wigner's reaction  $\hat K$-matrix were theoretically studied for chaotic systems with T-invariance either broken or not, and for arbitrary losses. The paper \cite{Fyodorov2020} generalizes the results of Ref. \cite{Fyodorov2015}  where  $T$-invariant and  $T$-violated systems in the limiting case of zero absorption were studied. The distribution of the modulus squared  of the off-diagonal elements  $|K_{ab}|^2$  for chaotic graphs with broken T-invariance
was derived  in Ref. \cite{Weaver2003}.

 Diagonal entries to the Wigner reaction matrix $\hat K$ have recently been observed for microwave networks with violated time-reversal invariance in the case of large losses \cite{Lawniczak2019b}. Contrary to the diagonal entries of $\hat K$, the off-diagonal elements of $\hat K$ have not been investigated experimentally yet.

The  distributions  of the off-diagonal elements $K_{ab}$ and $K_{ba}$ of the
$2\times 2$ reaction $\hat K$-matrix   can be
 obtained from the two-port scattering matrix  $\hat S$ of the network after elimination of direct processes, which are not chaotic \cite{Fyodorov2005,Hemmady2006}

\begin{equation}
\label{Eq:1}
\hat K =i\frac{\hat S-\hat I}{\hat S+\hat I}.
\end{equation}
The matrix $\hat I$ denotes the $2\times 2$ identity matrix.
   The
 matrix $\hat K$ is related to the  normalized impedance $\hat z$ \cite{Hemmady2006}: $\hat K=-i\hat z$ and is hermitian only without absorption.

In this article we consider  microwave networks with broken time-reversal invariance and characterized by intermediate and large width of the resonances $\gamma > 5.39$.
For such large values conventional indicators for $T$-violation such as short- and long-range spectral correlation functions \cite{Mehta1990}, the nearest neighbor level spacing distribution or the spectral rigidity are very difficult, if not impossible to use because individual levels are hardly distinguishable. Therefore, we will bypass this severe problem using the enhancement factor $W$ \cite{Lawniczak2010} as a probe of broken time-reversal.

Microwave networks consist of vertices (microwave joints) connected by
 edges, realized by coaxial SMA-RG402 cables.   The SMA-RG402 coaxial cable contains a center conductor of radius $r_1$ = 0.05 cm surrounded by a Teflon insulating layer having a
dielectric constant  $\varepsilon \simeq 2.06$ \cite{Savytskyy2001}.
The insulating layer is surrounded by a tubular conductor of
radius $r_2$ = 0.15 cm.
 Inside a coaxial cable and below the cut-off frequency of the  TE$_{11}$ mode  only the fundamental transverse-electromagnetic (TEM) mode can  propagate.
  The TE$_{11}$ mode cut-off frequency for the SMA-RG402 coaxial cable is $\nu_{cut} \simeq \frac{c}{\pi (r_1+r_2)
 \sqrt{\varepsilon}} \simeq 33$ GHz \cite{Jones}, where $c$ is the speed of light in the vacuum. Absorption of the networks was
 controlled by changing the total lengths of the networks and adding to the networks microwave 1~dB and 3~dB attenuators.

\begin{figure}[h!]
\includegraphics[width=1.0\linewidth]{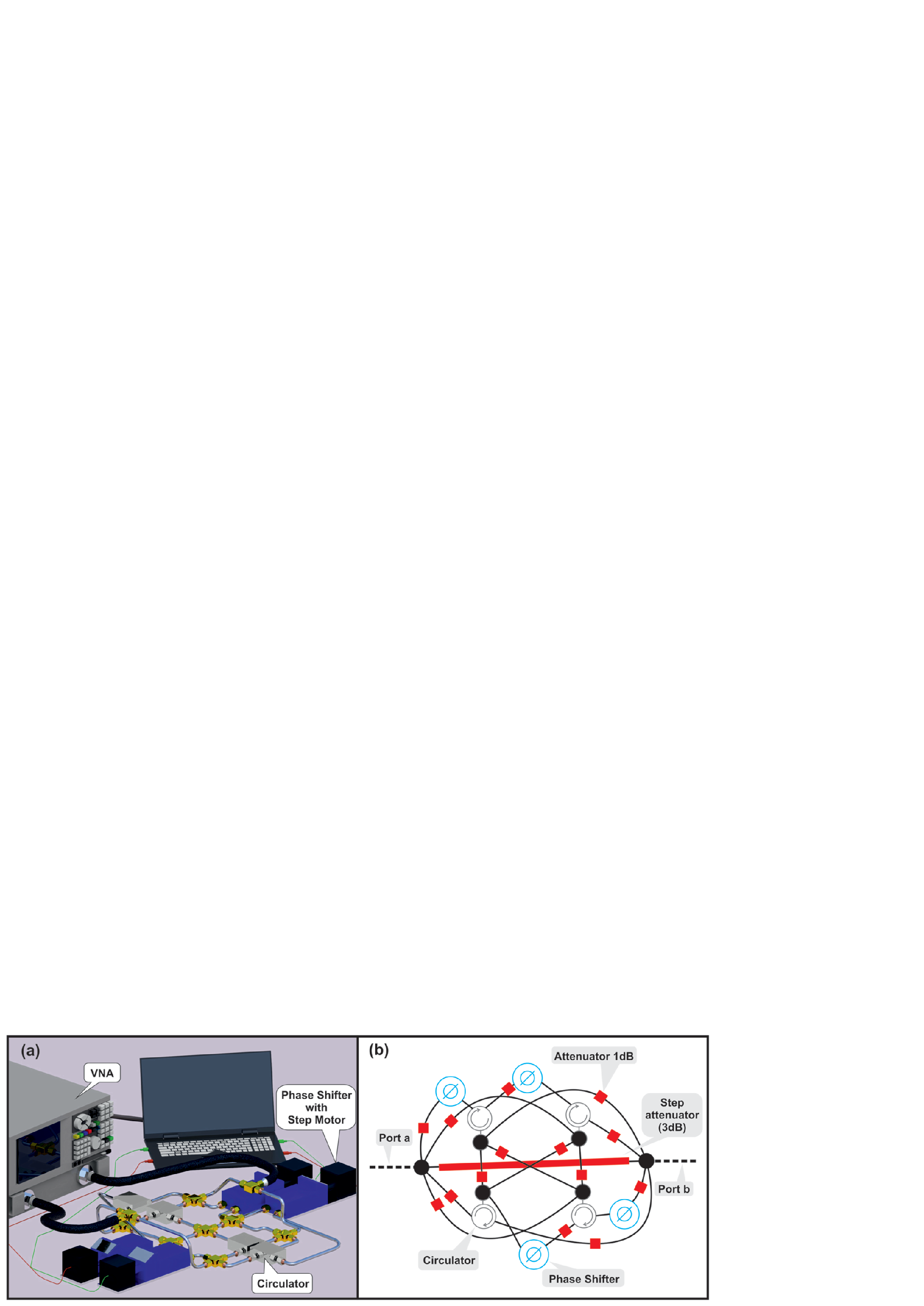}
\caption{(a) The experimental setup for
measuring the scattering matrix $\hat S$ of the microwave networks with violated $T$-invariance and intermediate value of the parameter $\gamma$.  The network consists nine microwave joints. Different realizations of the network were obtained using four phase shifters. The $T$-violation was induced with four Anritsu PE8403 microwave circulators. (b) The experimental setup for
measuring the scattering matrix $\hat S$ of the hexagon  microwave networks with violated $T$-invariance and large parameter $\gamma$. Also here, different realizations of the network were obtained using four phase shifters and  $T$-violation was induced with four Anritsu PE8403 microwave circulators. The matrix $\hat S$ was measured at the inputs of the 6-joint vertices.
}\label{Fig1}
\end{figure}

The two-port scattering matrix $\hat S$ of the 9-vertex and 6-vertex microwave networks
required for the evaluation of the Wigner's matrix $\hat K$ and the enhancement factor $W$ was measured using the setups shown in Fig.~1(a) and Fig.~1(b), respectively.
The $T$-violation was induced by four Anritsu PE8403 microwave Faraday circulators with low insertion loss which operate in the frequency range $3-7$~GHz. The microwave circulators are non-reciprocal three-port passive devices. A wave entering the circulator through port 1, 2 or 3 exits at port 2, 3, or 1, respectively, as shown schematically in Fig.~\ref{Fig1}(a).
  The ensembles of different microwave networks realizations were created by changing the lengths of four edges of the networks using the phase shifters visible  in Fig.~\ref{Fig1}(a) and Fig.~\ref{Fig1}(b) in such a way that the total ``optical" lengths of the networks were preserved.
  The hexagon network shown in Fig.~1(b) was built to investigate networks with larger parameter $\gamma$ and with larger total ``optical" length than the 9-vertex network. To increase even further the internal absorption of the network in order to verify the theory \cite{Fyodorov2020} in the limit of large losses, 1~dB attenuators were used on its 14 edges. The direct processes on the edge connecting directly the two 6-joint vertices (ports $a$ and $b$ of the network) were minimized by applying a 3~dB step attenuator.
The total ``optical" length of the 9-vertex network including edges, phase shifters, joints, and circulators, was 361 cm while the total ``optical" length of the hexagon, 6-vertex network, including all previous components and 1~dB and 3~dB attenuators was 662 cm.
 The scattering matrix $\hat S$ of the networks was measured in the frequency
range $3-7$ GHz by a vector network analyzer (VNA), Agilent E8364B.  The microwave networks were connected to the VNA through  the leads - flexible
microwave cables HP 85133-616 and HP 85133-617.
The experimental results were obtained by averaging over 1500 and 983 realizations of the 9-vertex network without attenuators
and the hexagon network containing  1~dB and 3~dB attenuators, respectively.

The elastic enhancement factor $W$ \cite{Lawniczak2010,Fyodorov2005,Dietz2010,Kharkov2013,Zheng2006} was used to monitor $T$-symmetry of the investigated systems. It can be obtained from the two-port scattering matrix $\hat S$ using the following relation
\begin{equation}
\label{Eq:2}
W=\frac{\sqrt{\mbox{var}(S_{aa})\mbox{var}(S_{bb})}}{\mbox{var}(S_{ab})},
\end{equation}
where $\mbox{var}(S_{ab}) \equiv \langle |S_{ab}|^2 \rangle
-|\langle S_{ab} \rangle |^2$ denotes the variance of the
matrix element $S_{ab}$.
  For intermediate and large parameter $\gamma $   the
enhancement factor $W$ is predicted to depend
 weakly on its value but is very sensitive to the ensemble \cite{Fyodorov2005}, approaching for  $\gamma \rightarrow +\infty $ the limit of
$W=2/\beta$.

 The exact results for the  distributions of the real $P(x_1)$ and imaginary $P(x_2)$ parts of  $S_{ab}$ in the framework of the supersymmetry RMT were obtained by A. Nock {\it at al.} \cite{Guhr2014}. In the $\beta=2$ case the distributions are identical and are given by the formulas (55-56) in Ref. \cite{Guhr2014}. These formulas depend on the antenna transmission coefficients $T_a$ and $T_b$, where $T_{m}=1-|\langle S_{mm}\rangle |^2$ for $m=a, b$ \cite{Fyodorov2005,Dietz2010}, and the transmission coefficients of $M$ parasitic absorption and leak channels $T_c$. Using the Weisskopf estimate \cite{Dietz2010} the parameter $\gamma$ characterizing the distributions $P(x_i)$ can be found in our case from the formula $\gamma=T_a+T_b+\sum_{c=1}^{M} T_c$, where $M\gg 1$.

In Fig.~2(a) the experimental distributions of the real  $P(x_1)$ and imaginary  $P(x_2)$ parts of the off-diagonal element of the scattering matrix $S_{ab}$, black full circles and black triangles, respectively,  are shown for the microwave networks with broken $T$-invariance. The fit of theory to data results in  $\gamma  =5.39 \pm 0.20$, corresponding to weakly overlapping resonances, is shown as a green broken line. The obtained value of $\gamma=T_a + T_b + \sum_{c=1}^{M} T_c$ was fitted for $M=100$  identical effective open channels with the transmission coefficient $T_c$, yielding $T_c=0.0361$. The  antenna transmission coefficients $T_a = T_b = 0.89 \pm 0.05$ were calculated using the formula $T_{m}=1-|\langle S_{mm}\rangle |^2$, where $m=a, b$. We checked that the  value of $\gamma  =5.39 \pm 0.20$ for $M \geq 50$ does not depend on the number of effective channels $M$. It is clearly seen that the observed statistics deviate strongly from the Gaussian approximation (small blue circles).

In Fig.~2(b) we show the measured distributions $P(x_i)$ of the real (black full circles) and imaginary (black triangles) parts of  $S_{ab}$   compared to the theoretical ones (green broken line) obtained for  $\gamma  =27.18 \pm 0.90$, which approaches the Ericson regime of strongly overlapping resonances. The parameter $\gamma$ was fitted for  $M=100$ channels and  $T_a = T_b = 0.56 \pm 0.05$, yielding  $T_c=0.261$. In this case the Gaussian approximation  (small blue circles) is much closer to the theoretical and experimental results.
 For both weakly and strongly overlapping resonances, the experimental  results are in excellent agreement with the theoretical predictions confirming accurately the validity of the theory exploiting the Heidelberg approach \cite{Guhr2014}.

In Table~1  the enhancement factor $W$ of the
scattering matrix $\hat{S}$ of the microwave networks  measured
  for two experimental values of the parameter $\gamma$ is compared to the theoretical prediction $W_{th}$. The enhancement factor $W_{th}$ was calculated using Eq.~(\ref{Eq:2}) by applying the formulas  (19) in Ref. \cite{Dietz2010} for the variances of the scattering matrix elements $S_{ij}$. In the calculations we used the transmission coefficients  $T_a=0.89 \pm 0.05$ and $T_b=0.89 \pm 0.05$, and $T_a=0.56 \pm 0.05$ and $T_b=0.56 \pm 0.05$ for the networks with intermediate and large parameter $\gamma$, respectively.  Next, the internal absorption  parameter $\gamma_{int}$ of the networks was calculated from $\gamma_{int}=\gamma -T_a-T_b$, which is equivalent to the sum over the transmission coefficients $T_c$, $\gamma_{int}=\sum_{c=1}^{M} T_c$.
  The comparison of the experimental and theoretical results clearly shows that both networks have broken $T$-invariance.
\begin{table}[!hbt]
\centering

\begin{tabular}{|c|c|c|c|}
\hline
 $\gamma $ & $W$ & $W_{th}^{GUE}$ & $W_{th}^{GOE}$\\
 \hline
 $5.39 \pm 0.20$ & $1.23\pm 0.22$ & $1.28$ & $2.19$\\
 \hline
 $27.18 \pm 0.90$ & $1.13\pm 0.12$ & $1.07$ & $2.04$\\
 \hline
\end{tabular}

\mbox{}

Table 1: The experimental  enhancement factor $W$ of the microwave networks  equipped with Faraday circulators
 compared to the theoretical predictions $W_{th}^{GUE}$ and $W_{th}^{GOE}$ for $GUE$ and $GOE$ systems, respectively, for two experimental values of the parameter $\gamma$.
\end{table}

We now turn to the observation of the statistics of the off-diagonal elements of the Wigner's
$\hat K$ matrix. According to the work of Fedeli and Fyodorov  \cite{Fyodorov2020}, the joint probability density function associated with complex Hermitian GUE matrices $\hat H_N$ of size $N \times N$, in the limit $N \rightarrow \infty$,   is given by

\begin{equation}
\label{Eq:3}
P(\Re K_{ab},\Im K_{ab})=\frac{\alpha ^2}{\pi} \lim_{x \rightarrow 2\pi \rho(\lambda)\alpha}  D_x \frac{\exp( -\sqrt{x^2+ 4\alpha ^2 |K_{ab}|^2}) }{\sqrt{x^2 + 4\alpha ^2 |K_{ab}|^2}},
\end{equation}
where $\Re K_{ab}$ and $\Im K_{ab}$ are the real and imaginary parts of the off-diagonal element  $K_{ab}$ and the operator $  D_x = \sinh (x)(1+ \frac{d^2}{dx^2}) - 2 \cosh(x) \frac{d}{dx}$. Equation~(\ref{Eq:3}) is parameterized by the parameter $\alpha $ and the spectral density  given by the Wigner's semicircular law $\rho(\lambda)=\frac{1}{2\pi}\sqrt{4-\lambda^2}$.

 In the model considered in \cite{Fyodorov2020} the losses are taken into account by allowing the spectral parameter (energy) $\lambda $ to achieve an  imaginary part by replacing $\lambda \rightarrow \lambda + i\alpha/N$ equal for all $N$ modes.
The correspondence between the parameter $\alpha$ and the parameter $\gamma$  can be established using the fact that the losses in a physical system such as absorption and the effect of openness can equivalently be taken into account by a purely imaginary shift of the scattering energy $\epsilon \rightarrow \epsilon +\frac{i}{2}\Gamma$, where   $\Gamma$ is the average width of resonances $\epsilon$ \cite{Fyodorov2005}.

An estimate for the value for $\alpha$ of our network can be determined by comparison of the rescaled widths of the spectral parameter $\frac{\alpha}{N\Delta_m}$ and resonances $\frac{\Gamma}{2\Delta}$, which leads to a simple relationship between $\gamma$ and $\alpha$,
\begin{equation}
\label{Eq:4}
  \gamma = 2\pi\Gamma/\Delta = 4\pi\rho(\lambda)\alpha,
 \end{equation}
 where $\Delta_m =1/(N\rho(\lambda))$ is the mean level spacing of GUE eigenvalues.

The probability density function  for the variables $u_1=\Re K_{ab}$ and $u_2=\Im K_{ab}$ can be  numerically evaluated upon performing the integral
\begin{equation}
\label{Eq:5}
P(u_i) = \int_{-\infty}^{+\infty}P(u_1,u_2)du_i.
\end{equation}

 The distributions $P(u_i)$  as well as  Eq.~(\ref{Eq:3}) depend on the parameters $x$ and $\alpha$.
However, in the limit $N \rightarrow \infty$ it is sufficient to stay in the center of the semicircle \cite{Guhr1998} where the spectral density of GUE eigenvalues is given by $\rho(0)=1/\pi$.

Hence, using Eq.~(\ref{Eq:4}) we can express the parameters $x$ and $\alpha$ in Eq.~(\ref{Eq:3}) as  $x=\gamma/2$ and $\alpha=\gamma/4$. In this way the theoretical distributions given by Eq.~(\ref{Eq:5}) can be compared directly to the experimental distributions $P(u_i)$  of the real and imaginary parts of the off-diagonal entries of the Wigner's $\hat K$-matrix.

In Fig.~3(a) the experimental distributions $P(u_1)$ and $P(u_2)$  of the real and imaginary parts of the off-diagonal element $K_{ab}$ of the Wigner's reaction $\hat K$ matrix, denoted by black full circles and black triangles, respectively,  are shown for the microwave networks with broken $T$-invariance.

 The theoretical distributions of the real and imaginary parts of the off-diagonal element $K_{ab}$ of the Wigner's reaction $\hat K$ matrix are marked by red broken line. The error marks indicate the errors arising from the $\gamma =5.39 \pm 0.20$ uncertainty. The experimental and theoretical distributions are in very good agreement. Furthermore, in agreement with theory \cite{Fyodorov2020} the experimental distributions $P(u_i)$  of the real and imaginary parts of the off-diagonal element $K_{ab}$ are very close to each other.
 Although, it is easily seen from Eq.~(\ref{Eq:3}) that the theory predicts that the both distributions are identical, the experimental confirmation of this property gives us an important validation of the experimental procedures. Fig.3(a) also shows that for weakly overlapping resonances the distributions $P(u_i)$ are significantly different from the Gaussian approximation (blue solid line).

The results for larger parameter $\gamma=27.18 \pm 0.90$ are shown in Fig.3(b). The experimental distributions $P(u_1)$ and $P(u_2)$ of the real and imaginary parts of the off-diagonal element $K_{ab}$ of the Wigner's reaction $\hat K$ matrix are denoted by black full circles and black triangles, respectively. The theoretical distribution is marked in Fig.3(b) with red broken line. The error bars mark the error caused  by the $\gamma =27.18 \pm 0.90$ uncertainty. It is clearly seen that this distribution is in very good agreement with the one we have observed. Because the large parameter $\alpha=\gamma/4=6.80 \pm 0.23$ the Gaussian approximation (small blue circles) is much closer to the experimental and theoretical distributions  $P(u_i)$ than in the case of much smaller parameter $\alpha=1.35 \pm 0.05$, where a big discrepancy between the experiment and the Gaussian approximation was observed.

In summary, we have measured the distributions  of the real and imaginary parts of the off-diagonal element of the scattering matrix $S_{ab}$  for chaotic networks with broken time reversal symmetry and for different losses. The experimental results are in very good agreement with the theoretical predictions from supersymmetry random matrix theory \cite{Guhr2014}.
We also experimentally determined the
distributions of the real and imaginary parts of the off-diagonal element $K_{ab}$ of the Wigner's reaction $\hat K$ matrix for chaotic networks with broken time reversal symmetry.   We show that  $K_{ab}$ is not Gaussian distributed when the levels do not completely overlap, in perfect agreement with recent theoretical prediction of Ref.~\cite{Fyodorov2020}.

Acknowledgments.  This work was supported in part by the National Science Centre, Poland, Grant No. UMO-2016/23/B/ST2/03979. LS thanks CNRS for support, contract No. 874365. We would like to thank Yan Fyodorov and Szymon Bauch for useful discussions.

\pagebreak

\begin{figure}[h!]
\includegraphics[width=1.1\linewidth]{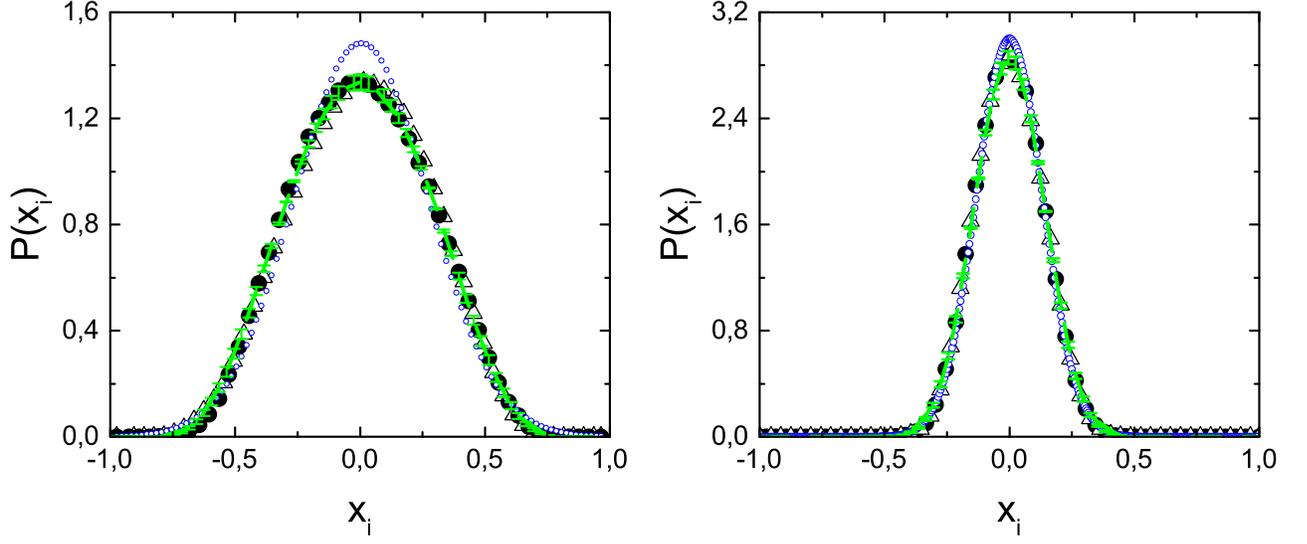}
\caption{(a) Experimentally evaluated distributions $P(x_1)$ and $P(x_2)$ of the real (black full circles) and imaginary (black triangles) parts of  $S_{ab}$ in the $\beta=2$ case at $\gamma  =5.39 \pm 0.20$. The theoretical fit with marked standard errors is denoted by green broken line.
The Gaussian approximation is marked with small blue circles. Panel (b) depicts the same quantities for $\gamma =27.18 \pm 0.90$.}\label{Fig2}
\end{figure}

\begin{figure}[h!]
\includegraphics[width=1.1\linewidth]{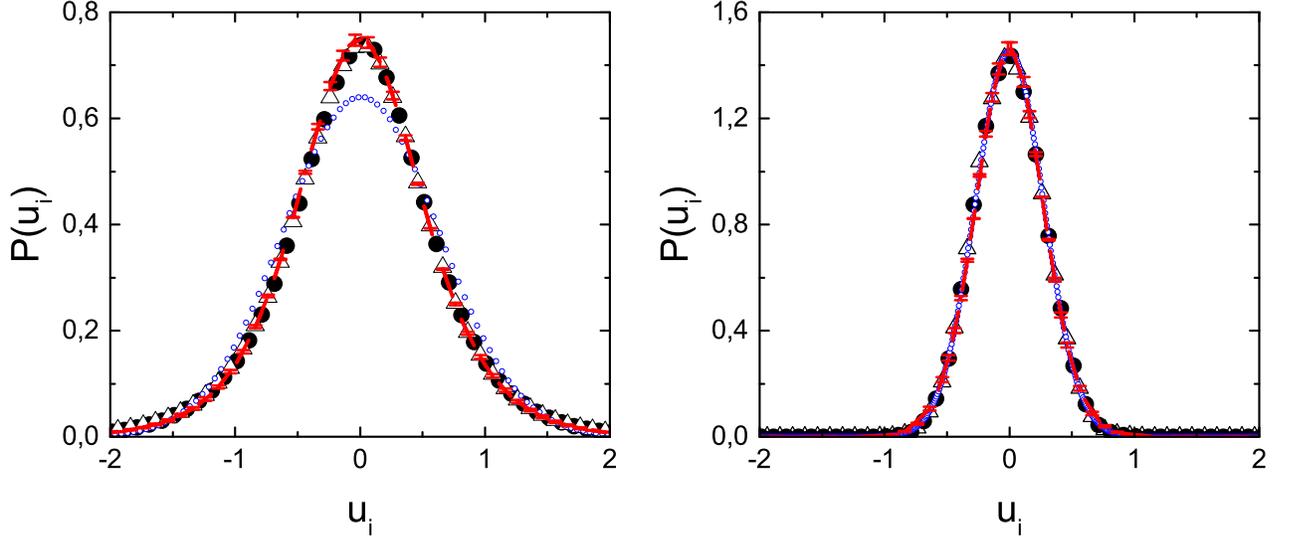}
\caption{(a) Experimentally evaluated distributions $P(u_1)$ and $P(u_2)$  of the real (black full circles) and imaginary (black triangles) parts of  $K_{ab}$ in the $\beta=2$ case at $\alpha=\gamma/4=1.35 \pm 0.05$ and $x=\gamma/2 =2.70 \pm 0.10$ . The theoretical results are denoted by red
 broken line. The error bars mark the errors caused  by the $\gamma = 5.39 \pm 0.20$ uncertainty.  The Gaussian approximation is marked with small blue circles. Panel (b) depicts the same quantities for  $\alpha=\gamma/4=6.80 \pm 0.23$ and $x=\gamma/2 = 13.59 \pm 0.45$. The error bars mark the errors caused  by the $\gamma =27.18 \pm 0.90$ uncertainty while the Gaussian approximation is marked with small blue circles.}\label{Fig3}
\end{figure}

\end{document}